\documentclass[a4paper,10pt,twoside]{cpc-hepnp}

\usepackage{multicol}
\usepackage{graphicx}
\usepackage{epstopdf}
\usepackage{booktabs}
\usepackage{amssymb,bm,mathrsfs,bbm,amscd}
\usepackage[tbtags]{amsmath}
\usepackage{lastpage}

\begin{document}

\fancyhead[c]{\small Chinese Physics C~~~Vol. XX, No. X (2017) xxx} \fancyfoot[C]{\small xxxx-\thepage}

\footnotetext[0]{Received 29 September 2016}

\title{Offline software for the DAMPE experiment}

\author{%
      Chi Wang()$^{1,2,*\email{chiwang@mail.ustc.edu.cn}}$
\and Dong Liu()$^{1,2}$
\and Yifeng Wei()$^{1,2,*\email{weiyf@ustc.edu.cn}}$
\and Zhiyong Zhang()$^{1,2}$
\and Yunlong Zhang()$^{1,2}$
\and Xiaolian Wang()$^{1,2}$
\and Zizong Xu()$^{1,2}$
\and Guangshun Huang()$^{1,2*\email{hgs@ustc.edu.cn}}$
\and Andrii Tykhonov$^{3}$
\and Xin Wu()$^{3}$
\and Jingjing Zang()$^{4}$
\and Yang Liu()$^{4}$
\and Wei Jiang()$^{4}$
\and Sicheng Wen()$^{4}$
\and Jian Wu()$^{4}$
\and Jin Chang()$^{4}$
}
\maketitle

\address{%
$^1$ State Key Laboratory of Particle Detection and Electronics, University of Science and Technology of China, Hefei 230026, China\\
$^2$ Department of Modern Physics, University of Science and Technology of China, Hefei 230026, China\\
$^3$ DPNC, UniversitšŠ de Genššve, CH-1211 Genššve 4, Switzerland\\
$^4$ Purple Mountain Observatory, Chinese Academy of Sciences, Nanjing 210008, China\\
}

\begin{abstract}
A software system has been developed for the DArk Matter Particle Explorer (DAMPE) mission, a satellite-based experiment. The DAMPE software is mainly written in C++ and steered using Python script. This article presents an overview of the DAMPE offline software, including the major architecture design and specific implementation for simulation, calibration and reconstruction. The whole system has been successfully applied to DAMPE data analysis, based on which some results from simulation and beam test experiments are obtained and presented.
\end{abstract}



\footnotetext[0]{\hspace*{-3mm}\raisebox{0.3ex}{$\scriptstyle\copyright$}2013
Chinese Physical Society and the Institute of High Energy Physics
of the Chinese Academy of Sciences and the Institute
of Modern Physics of the Chinese Academy of Sciences and IOP Publishing Ltd}%

\begin{multicols}{2} 

\section{Introduction}

The DArk Matter Particle Explorer (DAMPE) is a satellite-borne detector for detection of high energy cosmic rays 
\footnote{J. Chang et al, ``The DArk Matter Particle Explorer mission", under review}.
It was launched on 17 December 2015 \cite{lab2_news} into an elliptical and sun-synchronous orbit at an altitude of 500 km. Its mission is foreseen to operate at least three years. The primary scientific objective of DAMPE is to indirectly search for dark matter (DM) by measuring the spectra of photons, electrons, and positrons ranging from 5 GeV to 10 TeV with high energy resolution ( $<$1.5\% at 800 GeV).

The DAMPE detector is composed of four subdetectors, as shown in Fig.~\ref{fig1}. The top is a double-layer plastic scintillator strip detector (PSD), which serves as anticoincidence for gamma detection and charge measurement for ions. The PSD is followed by a silicon-tungsten tracker (STK), which is made of six silicon micro-strip layers with three 1.0-mm-thick tungsten plates inserted in front of tracking layer 2, 3, and 4. The BGO electromagnetic calorimeter (BGO ECAL) in the middle is composed of 308 BGO crystal bars with a dimension of 2.5 $\times$ 2.5 $\times$ 60.0 cm, which is $\sim$31 radiation lengths, making it the deepest calorimeter used in space experiments so far. At the bottom, a neutron detector (NUD) layer is added to improve the electron/proton separation capacity.

\begin{center}
\includegraphics[width=9.00cm]{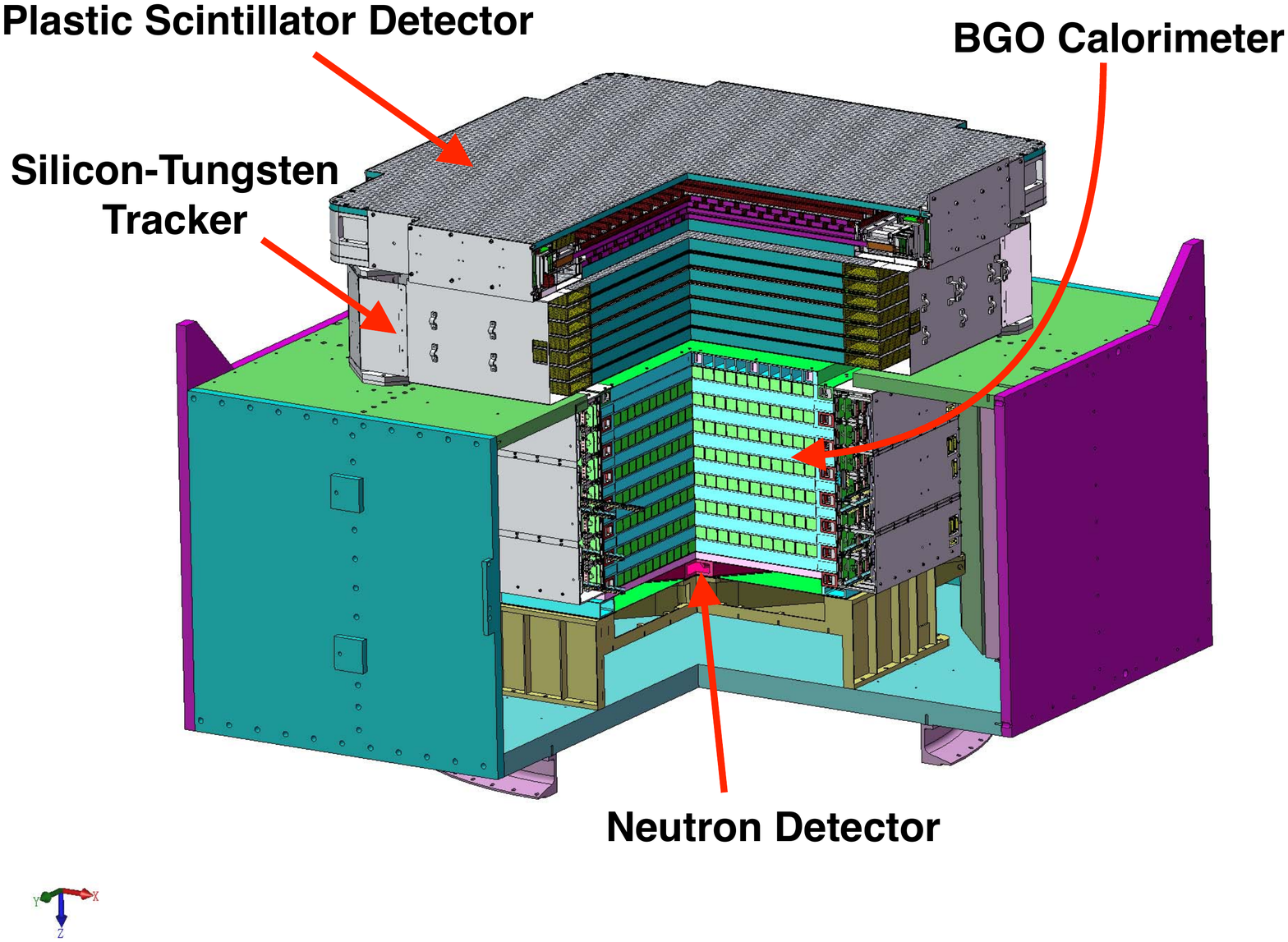}
\figcaption{\label{fig1} Structure of the DAMPE detector.}
\end{center}

Optimum operation of DAMPE requires an immense amount of hardware development effort, and also, to achieve the final physical goals, it is essential to develop offline software that is suitable for the special experimental data pipe.


This article first gives a review of the general requirements for the DAMPE software. Then, the basic software architecture is described and its implementation is introduced. Finally, some applications for DAMPE orbit calibration and beam test analysis are also presented.

\section{Requirements}

The main features of high-energy physics (HEP) experiments, such as those at the Large Hadron Collider (LHC), are the complexity of the detector system and the wide distribution of participants. Given the efficiency of worldwide communication among physicists doing analysis work, non-uniform programming logic from individual developers is inadequate in group collaboration. Besides, experiments are expected to run many years and therefore changes in analysis technologies must be envisaged. Thus it is necessary to develop flexible software that can withstand the transition, have the capability to integrate efforts from individual persons smoothly and be easily maintained over the long timescale of the project.

The idea of using uniform software is generally accepted in HEP experiments. GAUDI \cite{lab3_Gaudi} is a well-designed framework driven by LHCb \cite{lab4_LHCb}. It has been adopted in many other projects, e.g., ATLAS Athena \cite{lab5_Athena}, HARP Gaudino \cite{lab6_Gaudino}, and BESIII BOSS \cite{lab7_Boss}.
 SNiPER \cite{lab8_Sniper} is another analysis software framework designed for noncollider experiments; it is being used by the Jiangmen Underground Neutron Observatory (JUNO) \cite{lab9_JUNO} and the Large High Altitude Air Shower Observatory (LHAASO) \cite{lab10_LHAASO}.

The Alpha Magnetic Spectrometer (AMS-02) is a high-energy particle detector designed to study the origin and nature of cosmic rays up to a few TeV in space \cite{lab11_AMS02}. It does not use an analysis framework like GAUDI, while the basic component for analysis is a series of specified event classes that are inherited from ROOT's \emph{TObject}. The Fermi Large Area Telescope (Fermi-LAT), the primary instrument on the Fermi Gamma-ray Space Telescope (Fermi) mission, is an imaging, wide-field-of-view, high-energy $\gamma$-ray telescope, covering the energy range from 20 MeV to $>$300 GeV. The data analysis software of Fermi-LAT has been designed within the framework of the HEADAS FTOOLS methodology, to ensure cross-mission compatibilities wherever possible and to minimize the learning curve for users of other high-energy astrophysics mission data sets.



Since DAMPE consists of several subdetectors, it is not easy to combine nonuniform reconstruction and calibration codes from them. In the collaboration, many participants are astronomers, who are not so familiar with the computing mode used by particle physicists. The most important thing is that we have to provide DAMPE with suited analysis tools for worldwide researchers.
With these considerations, we started to design the DAMPE offline data analysis software system with an architecture that could be used to organize development activities.

With the experience of data analysis in HEP experiments, the first choice should be GAUDI. However, after 15 years of development, GAUDI has evolved into a bulky system and most of its complex functionalities are not necessary for DAMPE. Adapting GAUDI for DAMPE is not a practical solution, given the time of adaptation and the cost of maintenance in the future as well. Using SNiPER is also not practical as it was in the design and development stage at the beginning of the DAMPE project. Thus, under the condition of limited manpower and time, we started to construct new lightweight software for DAMPE.

The key challenge is how to provide a uniform development environment for programmers and a simple, flexible manipulation manner for the end users. Such a requirement is not only helpful for software development and maintenance but also significant for improving physics analysis quality and efficiency. The overall design of the DAMPE offline software takes into account the following principles:



\begin{itemize}
\item \textit{Separated environments for developer and user}:\quad One person may have different roles, as a software developer or a high-level user, at different times. It requires multiple computing environments. During development, developers could substitute their own versions of certain program modules while still using the standard version of the rest. After the developing task is done, the certain module could be easily integrated into the software system and become a new standard version.
\end{itemize}

\begin{itemize}
\item \textit{Data separation}:\quad The analysis work is a data-driven application, so the overall design should distinguish data objects from algorithm objects. In general, methods used to process data objects of some types and produce new data objects of other types have a much higher possibility of being modified than the data objects. Data and algorithms should be decoupled into two modules in the software. Besides, all different types of data should be accessible in an uniform manner and selected dynamically.
\end{itemize}


\begin{itemize}
\item \textit{Ease of use}:\quad The software system must be easily used by collaboration physicists who are not computing experts and cannot devote a lot of time to learn computing techniques. Two aspects must be addressed: For developers, physicists could provide a series of algorithms with different analysis technologies while the software could help them focus on implementation of the core functionality without being disturbed by general minor issues, such as data communication among modules. For end users, the software should be flexible enough to arrange the analysis chain, and adding or replacing certain modules should be possible by just changing a few lines in a steering file at run time.
\end{itemize}

Based on the above, C++ has been identified and adopted to build the main part of the software. By taking advantage of Boost Python \cite{lab12_boost}, the system has the capability to manipulate a specific task in Python script \cite{lab13_python} in an extremely easy and flexible manner. CINT \cite{lab14_cint} is an interpreter for C++ code and widely used in HEP analysis, and our software also has the capability of running in ROOT \cite{lab15_root} as a plug-in module.

\section{Basic Architecture}
The basis of the software is a pre-defined architecture that defines the fundamental components and the interface. The fundamental components of the software comprise five parts:

1. \textbf{\emph{Kernel}}, which is the steering part of the whole offline software.

2. \textbf{\emph{Event module}}, which contains all reconstruction information based on the ROOT input/output (I/O) stream, as well as detector simulation and digitization information.

3. \textbf{\emph{Algorithms}}, which are the central component of the event data processing.

4. \textbf{\emph{Event filters}}, which are used to reduce the input events into concrete algorithms.

5. \textbf{\emph{Services}}, which aim at providing common functionality needed by the algorithms.

The architecture of the software is shown as the object diagram in Fig.~\ref{fig2}.

\end{multicols}
\begin{center}
\includegraphics[width=10cm]{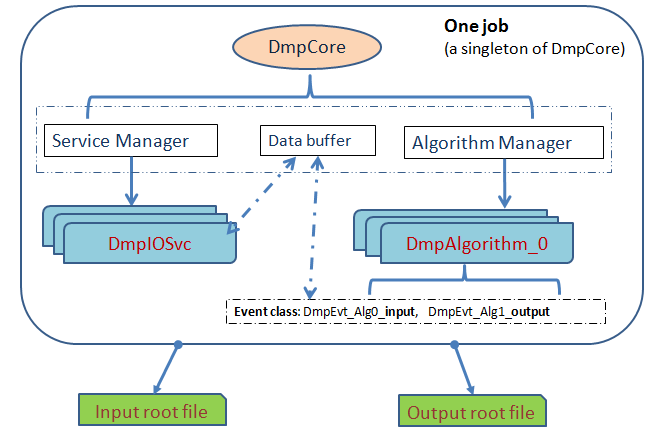}
\figcaption{\label{fig2} DAMPE software framework object diagram.}
\end{center}
\begin{multicols}{2}

\subsection{Kernel}
The kernel is the specification of a piece of software, and the DAMPE software kernel is composed of three parts: \emph{DmpCore}, \emph{DmpServerManager}, and \emph{DmpAlgorithmManager}. The main function of \emph{DmpCore} is to drive a job, including initialization, event loop and finalization.

As shown in Fig. \ref{fig2}, a job is a singleton of the top-level manager (\emph{DmpCore}), which manipulates the algorithm manager and service manager. The software provides a data communication mechanism and execution logic for applications. Meanwhile, it transfers the arrangement and deploys the proper concrete algorithms to users at run time via a Python script. With a predefined interface, it is possible to replace a concrete component with another one, but others remain unchanged. 

\subsection{Event module}
The event module holds the description of the event data items that are exchanged between algorithms in an application. It is the central part of an experiment. The production flow of event data represents the analysis logic. Mostly, the developer does not have the full picture of the event type for the next stage, so expansibility of the event data module is required. Data taken on board by DAMPE consist of raw scientific events, which are directly read out from subdetectors, 
and satellite status data, such as temperatures and position and velocity of the satellite, which are periodically sampled with different time scales. Different data are described with corresponding event classes. The first step of data processing is to convert raw binary orbit data into raw root format data. Then, the reconstruction algorithm is applied to create more physically meaning data, such as, for an example of BGO ECAL, from \emph{DmpEvtBgoRaw} (analog to digital convertor (ADC) counts of the photomultiplier tube (PMT)) to \emph{DmpEvtBgoHits} (energy of each BGO bar). The \emph{DmpEvtBgoHits} class is also used to record the output Monte Carlo information of the BGO subdetector. Figure \ref{figevtcls} shows the event classes for raw BGO calorimeter data containing dynode IDs \cite{MDY}, see minimum detection unit shown in Fig. \ref{fig8}, and ADC counts of PMTs, which are converted to hits data containing BGO bar IDs and the energy deposited in each bar.

\begin{center}
\includegraphics[width=8cm]{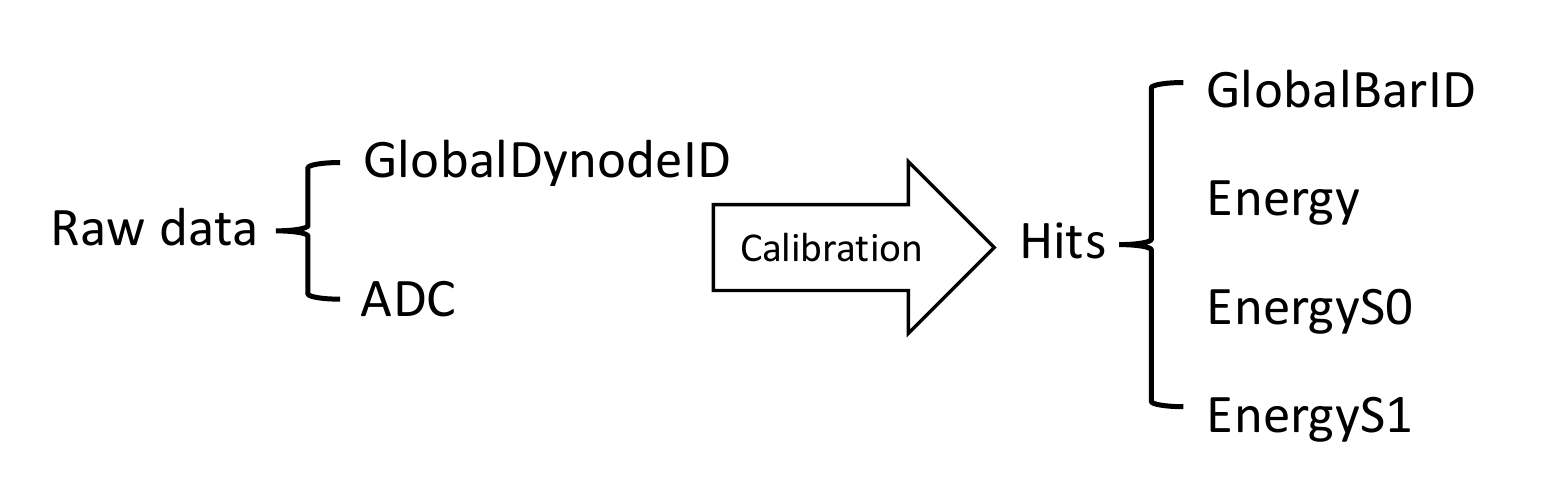}
\figcaption{\label{figevtcls} Event class content for raw data and hits data from the BGO calorimeter. Raw data contain ADC information and Hits contains energy deposited in each BGO bar, with "EnergyS0" and "EnergyS1" representing energy reconstructed on sides 0 and 1, respectively, of the bar.}
\end{center}

In the event module, all event classes inherit from \emph{TObject} of ROOT. Therefore, they naturally adopt the ROOT I/O stream and enable us to implement the data management mechanism by using the mediator pattern \cite{Pattern}.

\subsection{Algorithms}
The essence of the event data processing applications lies in the physics algorithms. Concrete algorithm classes must inherit from the base algorithm class (\emph{DmpVAlg}), which is provided by the software, to manipulate algorithms via the generic interfaces without knowing what they really do.

The core functionality of an algorithm is in producing event data one by one. This entails implementing a certain function \emph{DmpVAlg}::\emph{ProcessThisEvent}() under the architecture of DAMPE offline software. The remaining small tasks, for instance preparing a new event or communicating with each other, are already implemented in the software. The user should focus on extracting useful information from the data that is used in a new algorithm.

The algorithms are core parts of applications and all components are reusable; for example, the tracking algorithm can be used for alignment and acceptance analysis. If a series of algorithms forms an application, the correct arrangement order is required to archive the newest data from each preceding algorithm in a specified sequence. 

\subsection{Event filters}
Most of the time, a concrete algorithm only accepts events that could pass a set of conditions, and some algorithms share some common requirements. To make the codes clear and simplify it, we designed another module, composed of event filters that are decoupled from both algorithms and event classes. The provided application programming interface (API) is \emph{DmpVFilter}, and the essential part of an event filter is a Boolean function \emph{DmpVFilter}::\emph{Pass}(), which is only responsible for property checking of a certain type of input event. If the event does not suit those conditions, it will be discarded before invoking the core analysis part of algorithm. The execution logic is shown in Fig. ~\ref{fig3}.

\begin{center}
\includegraphics[width=8cm]{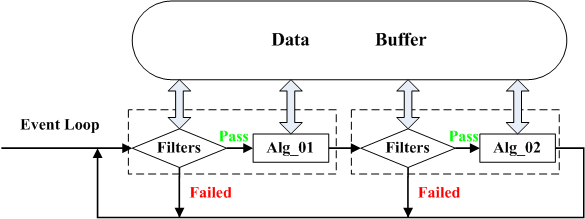}
\figcaption{\label{fig3} Execution logic of the algorithm with an event filter.}
\end{center}

This design is extremely useful for DAMPE data analysis. For instance, most subdetector calibration algorithms have to discard events that come from the South Atlantic Anomaly (SAA), and therefore the orbit filter (\emph{DmpFilterOrbit}) is developed. The input event class of the orbit filter is the event header that contains the time stamp of each event. By checking the gradient of the trigger rate at different times, we can distinguish whether or not the event is in the SAA. Besides, sometimes we need to compare analysis results in different sections of latitude and longitude, which is also provided by the orbit filter. All event filters can be constructed and inserted into any algorithm at run time dynamically and it makes DAMPE orbit data analysis more flexible and effective.


\subsection{Services}
The software involves many frequently used functions, which are implemented as services. The difference between a service and an algorithm is whether it reads or analyzes event data. Services are widely used by algorithms; e.g., an I/O service provides I/O support and an Extensible Markup Language (XML) service records all configuration values once a job starts. 

It is foreseen that more and more physicists will join the data analysis work of DAMPE in the future, so we designed the system with plug-in mechanisms, and the advantages of a modularized design will become more and more evident.

\section{Software implementation}

\subsection{Simulation}
The Monte Carlo (MC) simulation includes the detector description, detector response to cosmic rays, and signal digitization. The materials and geometry needed for the detector are read from Geometry Description Makeup Language (GDML) files, which are translated from AutoCAD files during the engineering design process. The constructed detector is shown in Fig. \ref{fig4}. 

The response of the detector to cosmic rays is handled by Geant4 \cite{lab16_G4}, a toolkit widely used in high-energy physics experiments to manage particle generation, propagation, and interactions. The whole simulation procedure is implemented using this software, which produces collections of energy hits for each sensitive detector element. 
 Compared to the common Geant4 application, the difference is that we do not build the simulation package as an executable file. We split the Geant4 Run Manager into three parts to satisfy the algorithm interface to let the software drive it. All Geant4 user actions and detector constructions are implemented in \emph{DmpVAlg}::\emph{Initialize}(). Since the event loop is already provided by the software, \emph{DmpVAlg}::\emph{ProcessThisEvent}() just invokes the functionality of one event execution of the Geant4 Run Manager. Then the resource is released in \emph{DmpVAlg}::\emph{Finalize}().

\begin{center}
\includegraphics[width=8.0cm]{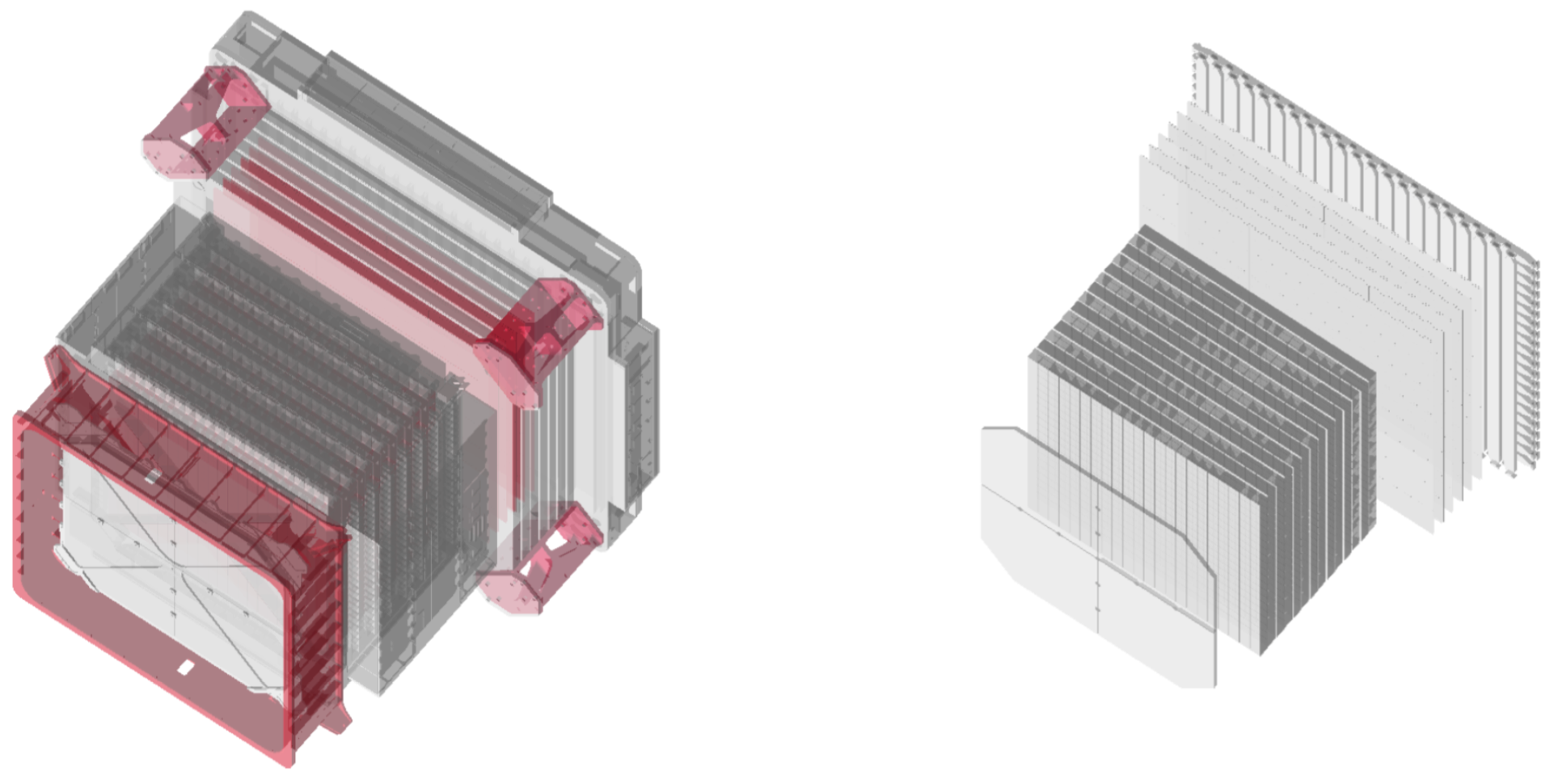}
\figcaption{\label{fig4} Constructed detector for the DAMPE experiment. (Left) Full detector. (Right) Sensitive part of the detector.}
\end{center}

In signal digitization, noise from the pedestal, PMT gains, and electronic fluctuations have to be taken in account. After digitization, MC data are stored in the same format as experimental data after energy reconstruction. Thus, the MC data can be processed with the same reconstruction algorithms and the simulation can provide an accurate representation of the instrument response for analysis. Also, for the orbit simulation, the same trigger conditions as used for real data have been implemented to simulate the final data stream. 
Figure \ref{fig5} shows the data flow of the DAMPE simulation. 

\begin{center}
\includegraphics[width=6.5cm]{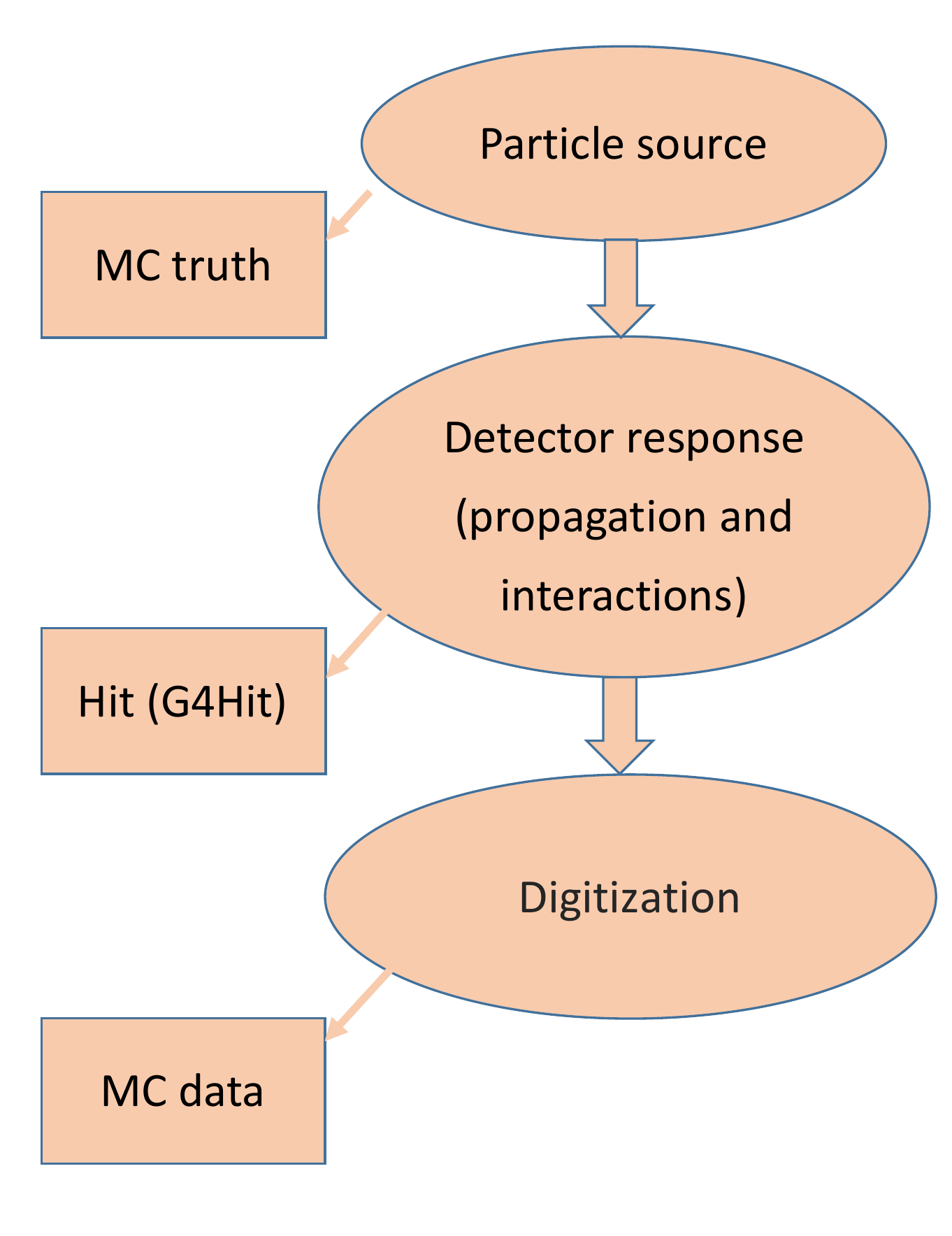}
\figcaption{\label{fig5} Data flow in the simulation. Rectangles show the data in the simulation.}
\end{center}

\subsection{Flight data processing}

In the DAMPE experiment, three data levels are defined:

\textbf{\emph{Level 0}}: binary data transmitted from the satellite, $\sim$12 GB per day.

\textbf{\emph{Level 1}}: 13 kinds of completed telemetry source packages, one for science data and 12 for housekeeping data, produced from level 0 data with operations such as data merging, overlap skipping and cyclic redundancy checking, etc.

\textbf{\emph{Level 2}}: science data product reconstructed with the DAMPE offline software. 

The data processing discussed here is for the period of level 1 translating to level 2. The processing pipeline of the DAMPE flight data includes raw data conversion (RDC) and calibration and reconstruction algorithms implemented in the DAMPE software. The RDC algorithm splits the level 1 science data into calibration files and observation files, and it then converts them into ROOT data files. Calibration files are used to extract calibration constants, which are used in the simulation and reconstruction algorithms. The reconstruction algorithm is used to reconstruct ADC counts of each subdetector to physical variables, such as energy and charge.

\subsubsection{Calibration}
The DAMPE software calibration package consists of calibration algorithms for the four subdetectors, which are independent of each other. Because of the modularized design of the software architecture, many common functions are extracted and packaged into algorithm libraries. All algorithms are created and appended to the algorithm manager one by one, which indicates the sequence of execution. A Python steering file starts the whole job. Here we present a calibration procedure to illustrate a concrete job running under the DAMPE software architecture. The job is to calibrate the absolute energy scale for the BGO ECAL with minimum ionizing particles (MIPs); the execution logic is shown in Fig.~\ref{fig6}. There are two filters inserted into the first algorithm: the trigger logical filter and the orbit filter; which are used to check the trigger logical tag and discard events in the SAA, respectively. After subtracting pedestal noise in the first algorithm, the event data (\emph{DmpEvtBgoRaw}) is updated and then transferred to the energy reconstruction algorithm, where the primary energy reconstruction is performed by using MIPs calibration constants and the ADC value of dynode 8 of the PMT, to generate new event data (\emph{DmpEvtBgoHits}). Then, the central filter, the MIPs selection filter, of the job starts to work based on the energy of each BGO bar. In the end, the MIPs spectrum is obtained and MIPs peak constants are calculated in the third algorithm.

All calibration constants are stored as ASCII files in the specified format. To manage these parameters easily, we developed a service, called \emph{DmpParHolder}, to  write and read calibration constants. Through timestamp of data, the \emph{DmpParHolder} can invoke the closest calibration parameters automatically in the period of reconstruction.

\begin{center}
\includegraphics[width=7.50cm]{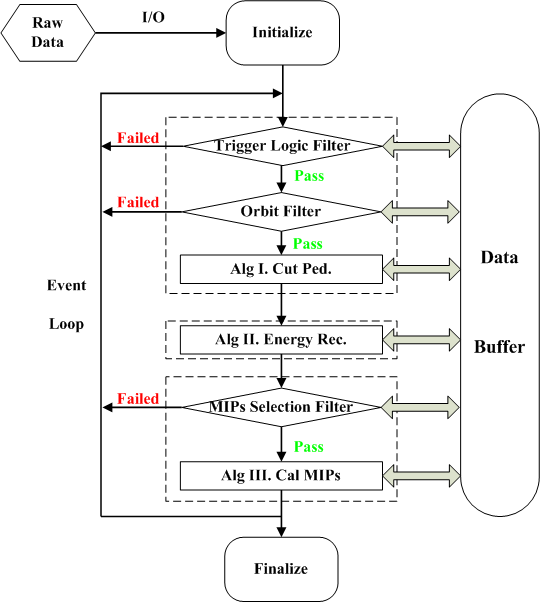}
\figcaption{\label{fig6} Execution logic of MIPs calibration application.}
\end{center}

\subsubsection{Reconstruction}
Data reconstruction is one of the central tasks of the offline data processing. For the DAMPE experiment, a complete chain of reconstruction algorithms has been developed and fitted into the software architecture. The reconstruction package contains the tracking reconstruction algorithm (BGO+STK), charge reconstruction algorithms (PSD+STK), and an energy reconstruction algorithm (BGO), etc. 

The key to event reconstruction for the DAMPE experiment is energy and tracking reconstruction. The most powerful instrument for tracking is the STK detector. Generally, the STK  track reconstruction algorithm gives a series of track candidates with different qualities. To obtain a high-precision track, when the direct reconstructed one obtained by the BGO ECAL is available, another global track finder algorithm combines the STK and BGO results and gives a global track. Meanwhile, since the scintillation light attenuates in the long BGO crystal (of 60 cm), the attenuation correction is applied in energy reconstruction, and its algorithm depends on the result of the tracking reconstruction. Therefore, we reconstruct a raw energy and global track first (version 0) and then an iteration reconstruction is implemented. The whole workflow is shown in Fig.~\ref{fig7}.

\begin{center}
\includegraphics[width=8.0cm]{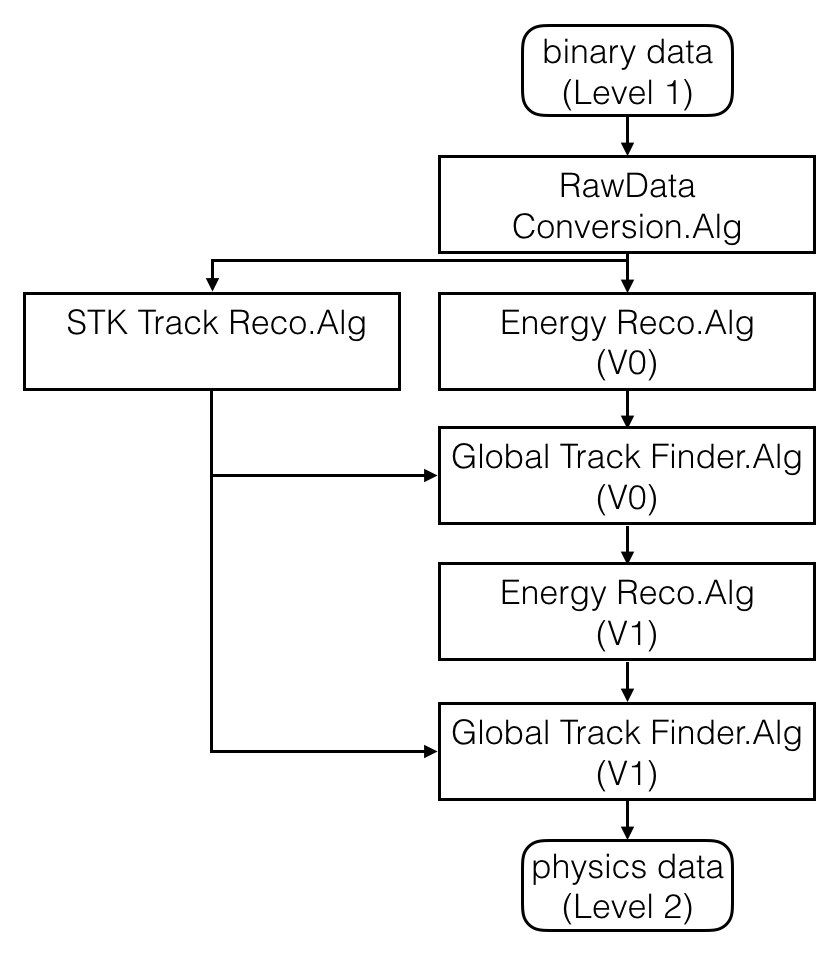}
\figcaption{\label{fig7} Workflow of energy and tracking reconstruction.}
\end{center}

\subsubsection{Housekeeping data}
For a satellite experiment, many engineering data, such as orbit latitude and longitude, are also essential in physical analysis. This specific information is called housekeeping data (HKD) in the DAMPE experiment.  The HKD packages are parsed from the satellite and then inserted into the housekeeping database after being processed by the ground support system (GSS). For the convenience of connecting HKD with physical analysis, an HKD service was developed in the DAMPE software. During the period of raw data conversion of science data, key HKD parameters needed in the analysis are stored into the ROOT files of science data through the interface provided by the HKD service. Users are also able to get these parameters from data with the HKD service.

\section{Application}

The DAMPE software have been verified and improved during ground test, including long-time cosmic ray test, thermal vacuum experiments, and beam test at CERN. We present some application results here.

The BGO ECAL comprises 14 layers of BGO crystals and each layer is composed of 22 BGO bars. One BGO bar is coupled with two PMTs on both its ends as shown in Fig.~\ref{fig8}. To extend the dynamic range, the signals are read out from three dynodes (Dy2, Dy5 and Dy8) of the PMT.

\begin{center}
\includegraphics[width=8cm]{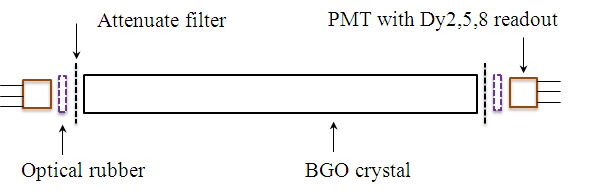}
\figcaption{\label{fig8} BGO ECAL minimum detection unit.}
\end{center}

The primary purpose of the BGO ECAL is to precisely measure the energy of incident particles. Since the energy scale obtained by the MIPs calibration is only available for Dy8, we perform the calibration of the gains of Dy8/Dy5 and Dy5/Dy2 before the energy reconstruction. For one event, we calculate a valid ADC value of Dy8 and transform it to energy by multiplying by the corresponding energy scale. When the readout ADC of Dy8 overflows, the valid ADC of Dy8 is calculated from the lower dynode and the calibration constants of the gains.

In the simulation, the BGO crystals are defined as the sensitive detector for the BGO ECAL. We record the total energy of each bar and simulate the light attenuation of BGO crystals via a parametric method using light attenuation calibration results \cite{lab17_BT}. The output data format of BGO ECAL is defined by the \emph{DmpEvtBgoHits} class. 



The reconstruction and simulation algorithms have been used for beam test data analysis. Figures~\ref{fig9} and \ref{fig10} show the energy linearity and energy resolution, respectively, of the DAMPE BGO calorimeter utilizing the electron beam test at CERN in 2014. Both simulation and beam test data indicate that the energy resolution is $<$0.8\% at 243 GeV, which is better than that of the design objective. More details about the beam test results can be found in Ref. \cite{lab17_BT}.

\begin{center}
\includegraphics[width=8.5cm]{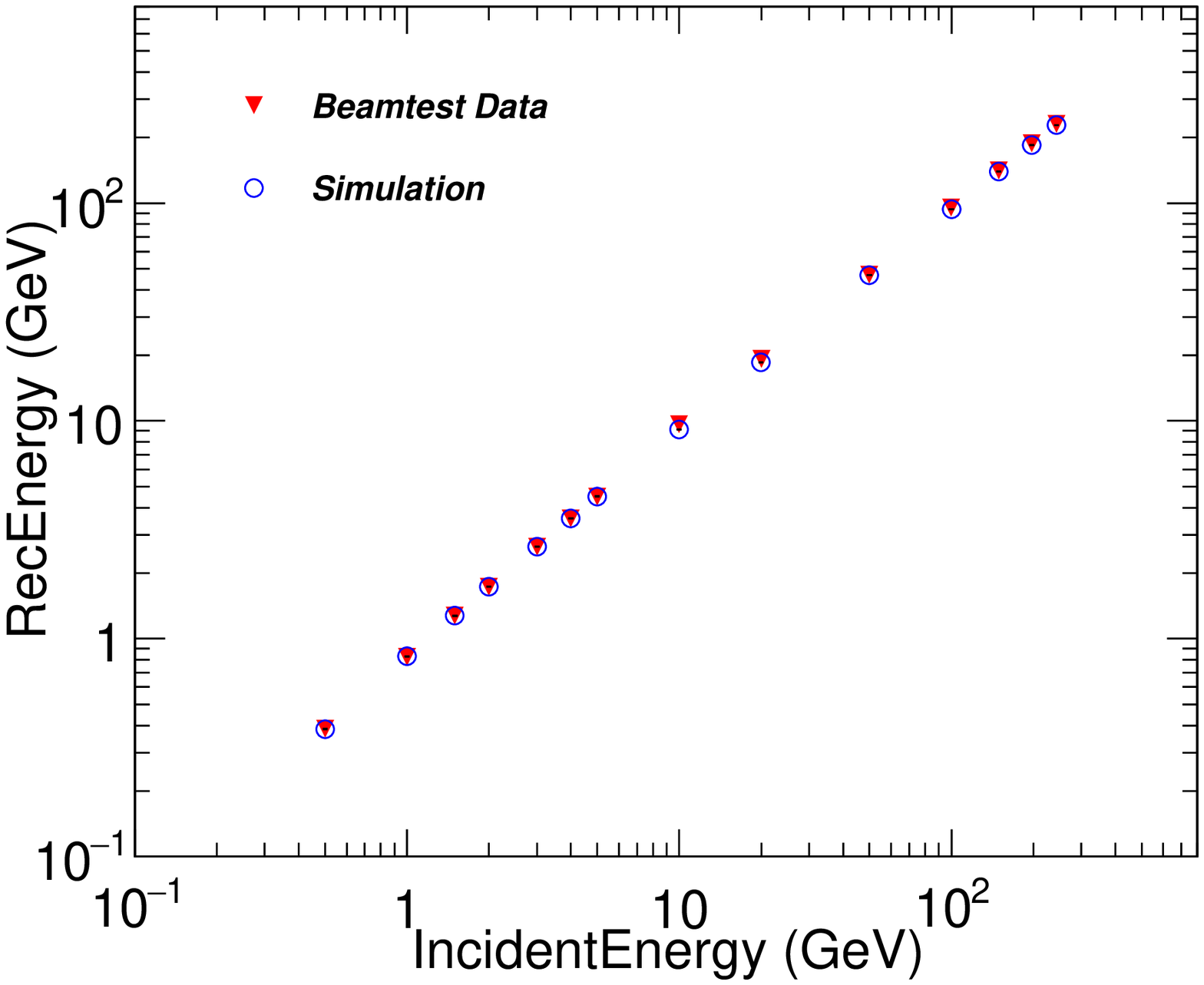}
\figcaption{\label{fig9} Energy linearity of the DAMPE BGO ECAL.}
\end{center}

\begin{center}
\includegraphics[width=8.5cm]{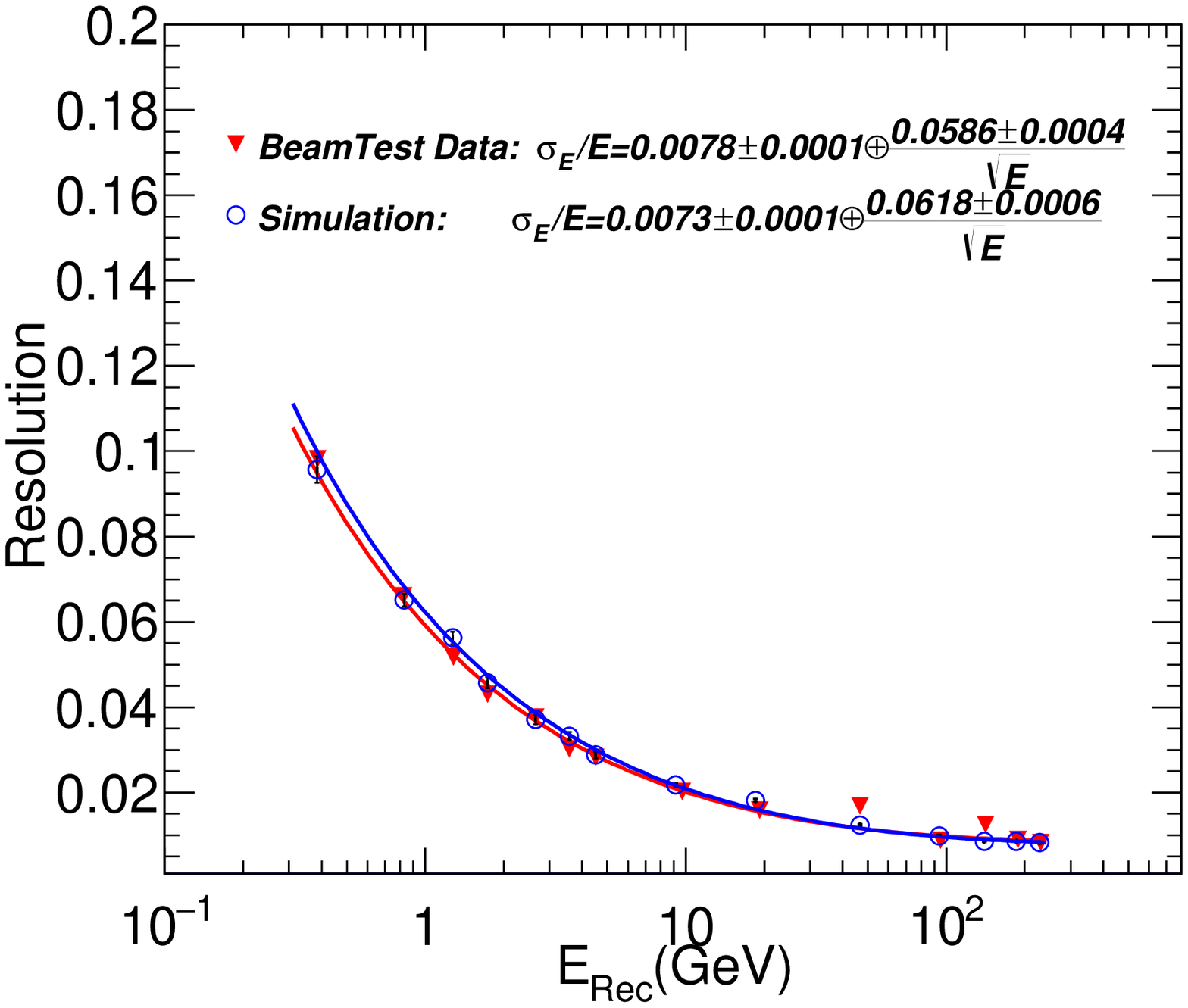}
\figcaption{\label{fig10} Energy resolution of the DAMPE BGO ECAL.}
\end{center}

DAMPE scientific data have been taken since 24 December 2015 after launch and the whole software system has been adapted for orbit data analysis. By inserting the MIPs selection filter into the energy reconstructed algorithm, we calculated the total deposited energy for orbit MIPs events, as shown in Fig.~\ref{fig11}. The black dots are orbit data; these are compared with the Monte Carlo simulation results (red histogram).


\begin{center}
\includegraphics[width=8.5cm]{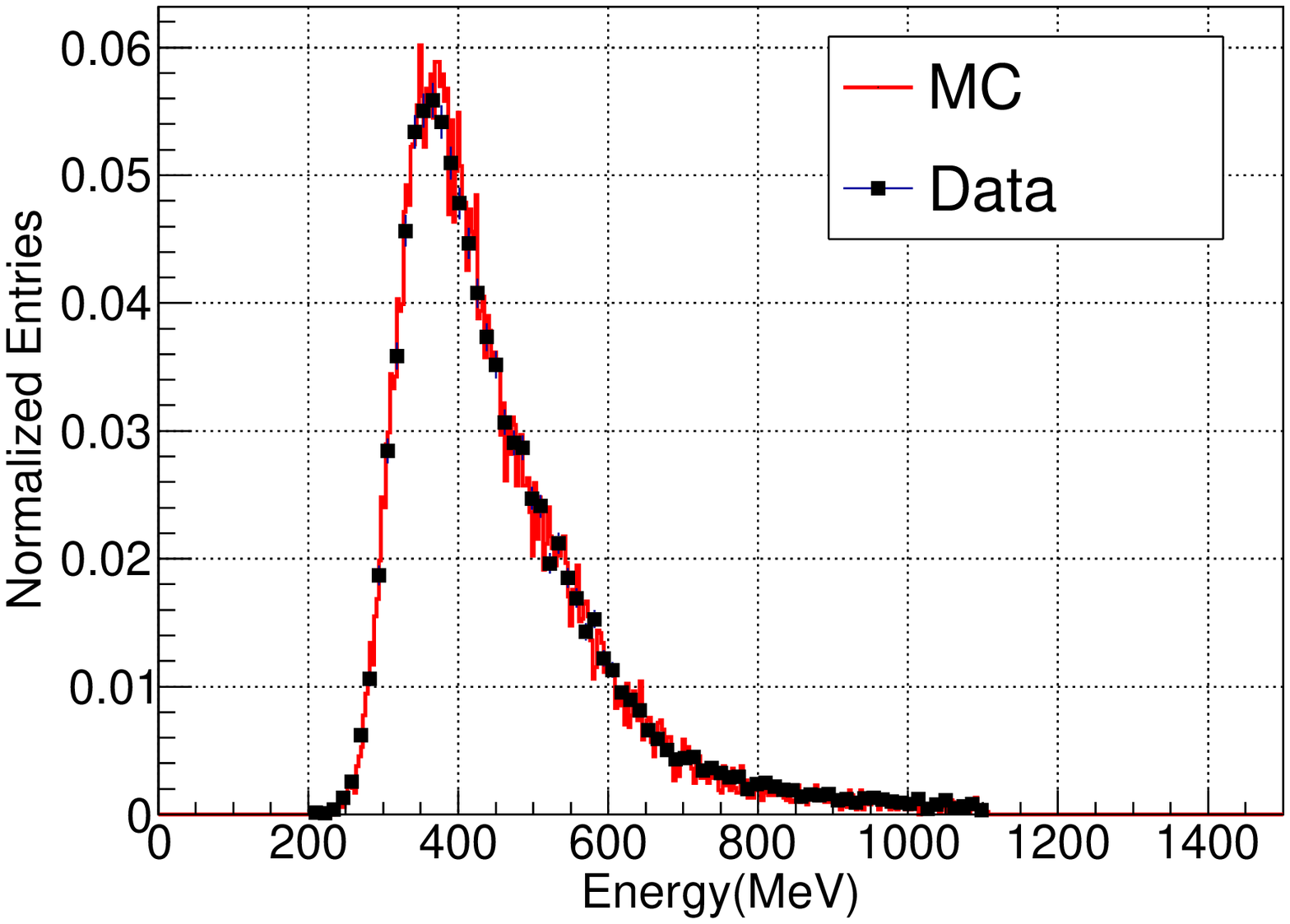}
\figcaption{\label{fig11} Reconstructed energy spectrum of proton MIPs.}
\end{center}

The good agreement between the simulation and real data confirms not only the validity of the DAMPE analysis software system but also the reliability of the DAMPE detector.

\section{Summary}
The software for the DAMPE experiment is user friendly, flexible and stable. It uses state-of-art techniques of computer science and satisfies most needs of DAMPE offline analysis. With the help of the architecture, collaboration-wide development works can be established in a more reliable and extensive manner. The system has been widely exercised during long-term cosmic test, three beam test experiments and has now been applied in orbit data analysis successfully. So far, the physics results have proved the effectiveness and robustness of the software. 

DAMPE is a pioneer of satellite scientific experiments in China, providing valuable experience for other experiments. The design principles for software developed for DAMPE presented in this paper can also be a useful guideline. The kernel and some classes with virtual functions can be reused to improve the development progress of software for new and similar experiments.

\vspace{4mm}
\subsection*{Acknowledgements}
This work was supported by the Chinese 973 Program, Grant No.2010CB833002, the Strategic Priority Research Program on Space Science of the Chinese Academy of Science (CAS), Grant No.XDA04040202-4, the Joint Research Fund in Astronomy under cooperative agreement between the National Natural Science Foundation of China (NSFC) and CAS, Grant No.U1531126, and 100 Talents Program of the Chinese Academy of Science.

The authors wish to thank Prof. Weidong Li and Dr. Tao Lin from Institute of High Energy Physics China for providing many beneficial suggestions on the software architecture.

%
\vspace{2mm}
\centerline{\rule{80mm}{0.1pt}}
\vspace{2mm}
%
%
%

\end{multicols}

\clearpage

\end{document}